# A new constitutive law for dense granular flows

Pierre Jop[1], Yoël Forterre[1] & Olivier Pouliquen[1]

*IUSTI, CNRS UMR 6595, Université de Provence, 5 rue Enrico Fermi, 13465 Marseille Cedex 13, France.*

**In many geophysical or industrial applications involving flows of grains, a continuum description would be of considerable help, for example in predicting natural hazards or in designing industrial processes. However, constitutive equations for dry granular flows are still matter of debates[1-10]. One difficulty is that grains can behave like a solid (in a sand pile), a liquid (when poured from a silo) or a gas (when strongly agitated)[11]. For the two extreme regimes, constitutive equations have been proposed based on kinetic theory for the collisional rapid flows[12], and soil mechanics for slow plastic flows[13]. The intermediate dense regime, where the granular material flows like a liquid, still lacks unified view and has motivated many studies the last decade[14]. The main characteristics of granular liquids are the existence of a yield criterion, i.e. the flow is not possible below a critical shear stress, and a complex shear rate dependence when flowing. In this sense, granular matter shares similarities with classical visco-plastic fluids like Bingham fluids. Inspired by this analogy and recent numerical[15,16] and experimental works [17-19], we propose here a new constitutive relation for dense granular flows. We then test our three-dimensional (3D) model by carrying out experiments of granular flows on a pile between rough sidewalls, where a complex 3D flow pattern develops. We show that, without any fit parameter, the model gives quantitative predictions for the flow shape and velocity profiles. These new results support the idea that a simple visco-plastic approach can capture quantitatively granular flows properties and could serve as a basic tool for modelling more complex flows in geophysical or industrial applications.**

Advances in our understanding of dense granular flows have been recently made by comparing different flow configurations[14]. The simplest configuration from a rheological point of view is the one sketched in Fig. 1 (inset). A granular material confined under a normal stress $P$ in between two rough planes is sheared at a given shear rate $\dot{\gamma}$ by applying a shear stress $\tau$. Da Cruz *et al.*[15] and Iordanoff *et al.*[16] have shown using dimensional arguments and numerical simulations, that for stiff particles the shear stress is proportional to the normal stress, with a coefficient of proportionality function of a single dimensionless number $I$ that they called the Inertial number:

$$\tau = \mu(I)P \quad \text{with} \quad I = \frac{\dot{\gamma}d}{\sqrt{P/\rho_s}} \tag{1}$$

where $\mu(I)$ is the friction coefficient, $d$ is the particle diameter and $\rho_s$ is the particle density. They found that the volume fraction $\Phi$ of the sample is also a function of $I$ but varies only slightly in the dense regime. The inertial number, which is the square root of the Savage number[20] or of the Coulomb number[21] introduced previously in the literature, can be interpreted as the ratio between two time scales, a macroscopic deformation time scale ($1/\dot{\gamma}$) and an inertial time scale $\sqrt{d^2\rho_s/P}$. By confronting results from the simple shear test with experimental measurements of granular flows on rough inclined planes[17,22], it can be shown that the friction coefficient $\mu(I)$ has the shape given in Fig. 1. It starts from a critical value $\mu_s$ at zero shear rate and converges to a limit value $\mu_2$ at high $I$. The following friction law can then be proposed which is compatible with the experiments[19]:

$$\mu(I) = \mu_s + \frac{\mu_2 - \mu_s}{I_0/I + 1} \tag{2}$$

where $I_0$ is a constant. Very recently, this simple description of granular flows has been successful in predicting two-dimensional configurations, capturing velocity profiles on inclined planes[14] and important features of flows on a pile[19]. However, the simple scalar law (Eq. 1) cannot be applied in more complex flows where shear in different directions is present and where a full three-dimensional rheology is needed.

We therefore propose the following 3D generalization of the friction law for a granular material. The basic assumption consists in neglecting the small variation of the volume fraction observed in the dense regime. The granular material is then described as an incompressible fluid with the internal stress tensor given by the following relations:

$$\sigma_{ij} = -P\delta_{ij} + \tau_{ij} \quad \text{and} \quad \tau_{ij} = \eta(|\dot{\gamma}|, P)\dot{\gamma}_{ij}$$

$$\text{with} \quad \eta(|\dot{\gamma}|, P) = \frac{\mu(I)P}{|\dot{\gamma}|} \quad \text{and} \quad I = \frac{|\dot{\gamma}|d}{\sqrt{P/\rho_s}} \quad (3)$$

where $\dot{\gamma}_{ij} = \partial u_i / \partial x_j + \partial u_j / \partial x_i$ is the strain rate tensor and $|\dot{\gamma}| = \sqrt{\frac{1}{2}\dot{\gamma}_{ij}\dot{\gamma}_{ij}}$ is the second invariant of $\dot{\gamma}_{ij}$. In this rheology, $P$ represents an isotropic pressure, and $\eta(|\dot{\gamma}|, P)$ is an effective viscosity, which definition is related to the friction coefficient $\mu(I)$ (Eq. 2). An important property of the proposed constitutive law is that the effective viscosity diverges to infinity when the shear rate goes to zero. This divergence insures that a yield criterion exists. Looking at Eq. 3 in the limit of $|\dot{\gamma}|$ going to zero, one can show that the material flows only if the following condition is satisfied:

$$|\tau| > \mu_s P \quad \text{where} \quad |\tau| = \sqrt{\frac{1}{2}\tau_{ij}\tau_{ij}} \quad (4)$$

The yield criterion takes then the form of a Drucker-Prager like criterion[23]. Below the threshold, the medium behaves locally as a rigid body. It is interesting to note that within this framework, the granular media can be viewed as a visco-plastic fluid[24]. The specificity compared to classical Bingham or Herschel-Bulkley fluids is that the effective viscosity depends both on the shear rate and on the local pressure. This property is linked to the frictional nature of stresses in granular media.

In order to test this rheology we performed experiments of granular flows on a heap confined between two lateral rough walls as sketched in Fig. 2. This configuration represents a severe test for the model, since it gathers in a single configuration several specifities of granular flows. First, when grains are released from the hopper, a steady regime is reached with a strongly sheared layer flowing on top of a static zone. The slope and the thickness of the flowing layer are selected by the system. Secondly, due to the rough sidewalls, a shear exists also in the transverse direction, the flow pattern being thus fully three-dimensional. The experiments are carried out using glass beads 0.53 mm in diameter. Our set-up, sketched in figure 2, consists in a channel partially closed at the bottom end to create a static pile on top of which the grains are flowing[19]. The sidewalls are made rough by gluing one layer of beads on them. The boundary conditions are then a zero velocity condition that induces strong shear in the lateral direction. The two control parameters of the system are the width $W$ of the channel and the flow rate per unit of width $Q$. The present study focuses on the steady and uniform regime characterized by a constant slope and a velocity aligned along the $x$-direction and invariant along the flow (a tiny $y$-component can be observed close to the wall, which remains 20 times smaller than the stream-wise velocity). In this steady uniform regime, we performed systematic measurements of the free-surface inclination $\theta$, of the free-surface-velocity profile $V_{surf}(y)$ using particle-imaging velocimetry and we get estimates of the thickness of the flowing layer $h(y)$ using an erosion method[19].

To compare the experimental results with the predictions of the local rheology, we perform numerical simulations of a granular fluid described by the constitutive law Eq. 3 and flowing in an inclined U-shaped channel with a no slip boundary condition at the three walls. The velocity $u(y,z)$ is assumed to be aligned with $x$ and to depend only on $y$ and $z$. To get the 3D steady velocity profile, we solve the incompressible Navier-Stockes equations with the internal stress being given by Eq. 3 using a finite different scheme. For the rheological parameters $\mu_s$, $\mu_2$ and $I_0$ coming into play in eq. 2, we choose the values given by the experimental data of flows on inclined plane[18] where the same particles were used (see Jop *et al.*[19] for the procedure how to compute this parameters): $\mu_s$=tan(20.9), $\mu_2$=tan(32.76) and $I_0$=0.279. This choice means that no fit parameter exists when we shall compare results from the simulations to the experimental data. A typical velocity profile obtained by the model is shown in figure 3. We first observe that a static zone develops at the base of the channel. The limit of the static zone varies across the channel, the flowing layer being larger in the centre than close to the walls. The second observation is that the velocity profile is truly three-dimensional and sheared in both *y* and *z* directions.

We have then tried to quantitatively compare the velocity profiles predicted by the simulations with the ones measured experimentally. In the simulation we impose the inclination and compute *a posteriori* the flow rate, whereas in the experiments, the flow rate is controlled and the inclination is measured. Figures 4a-c show the free-surface-velocity profiles obtained in the experiments and simulations for different widths and different flow rates. The experimental data are the symbols and the continuous lines are the prediction of the 3D rheology. The agreement is very good and quantitative. A slight deviation between experiment and model is observed in the narrower channel, 16.5 particle-diameters wide. In Fig 4d, we also compare the prediction of the theory for the thickness of the flowing layer. In both theory and experiments, the flowing layer is thicker in the centre than at the walls. A quantitative agreement is again observed, although the simulation systematically overestimates the flowing layer thickness. This could be due to the erosion method used for estimating the thickness measurement that is not very precise. All these results show that the proposed rheology gives quantitative predictions for this complex 3D flow, a striking success for a model with no fit parameter.

We have systematically carried out experiments for a wide range of flow rates and channel widths, and within 15% a quantitative agreement is always observed. To compare in a systematic way experiments and simulations, it is interesting to notice that simple scalings can be predicted from the rheology proposed. It is easy to show analytically that one can get rid of the width of the channel in the equations of motion by using the following dimensionless variables: $\tilde{z} = z/W$, $\tilde{y} = y/W$, $\tilde{V} = Vd/g^{1/2}W^{3/2}$ and $\tilde{Q} = Qd/g^{1/2}W^{5/2}$ (see supplementary information). It follows that the inclination of the pile $\theta$, the maximum velocity in the centreline of the channel $\tilde{V}_{max}$, and the maximum flowing thickness $\tilde{h}_{max}$ should depend only on $\tilde{Q}$. In figure 5, we show that the experimental measurements follow the predicted scaling and that the numerical simulations (continuous lines in Fig. 5) give quantitative predictions. One interesting result of this scaling analysis is that the thickness $h$ scales with the width $W$ meaning that the thickness of the flowing layer is not an intrinsic property of the granular media but is controlled by the width of the channel and the flow rate.

From this study, we can conclude that the simple visco-plastic constitutive law proposed seems very successful to describe dense granular flows. Without any fit parameter, it quantitatively captures the complex three-dimensional sheared flow observed when grains flow in between two rough walls. Limits of the approach exist, which mainly concerns the yield criterion. Within the proposed constitutive law, the flow threshold is simply described by a Coulomb criterion. However, the transition between solid and liquid like behaviour in granular matter seems much more complex, involving shear bands[25,26], intermittent flows[27] and hysteretic phenomena[28-30]. Such features still have to be included in a more comprehensive rheology. However, we believe that the simple visco-plastic rheology presented in this study represents a minimal model that quantitatively captures the basic features of granular flows important in many applications. We think it can serve as a base for further development to take into account more accurately the complex yield features specific to granular matter.

**Supplementary Information accompanies** the paper on **www.nature.com/nature**.

**Competing interests statement** The authors declare that they have no competing financial interests.

**Correspondence** and requests for materials should be addressed to P.J. (e-mail: Pierre.Jop@polytech.univ-mrs.fr).

References

1. Savage, S.B. Analysis of slow high-concentration flows of granular materials. *J. Fluid Mech.* **377,** 26 (1998).

2. Mills, P., Loggia, D. & Texier, M. Model for stationnary dense granular flow along an inclined wall. *Europhys. Lett.* **45**, 733 (1999).

3. Aranson, I. S. & Tsimring, L. S. Continuum description of avalanches in granular media, *Phys. Rev. E,* **64**, 020301 (2001).

4. Pouliquen, O., Forterre, Y. & Le Dizes, S. Slow dense granular flows as a self-induced process. *Adv. Compl. Syst.* **4**, 441–450 (2001).

5. Bocquet, L., Losert, W., Schalk D., Lubensky T. C. & Gollub, J. P. Granular shear flow dynamics and forces: Experiment and continuous theory. *Phys. Rev. E* **65**, 011307 (2002).

6. Ertas, D. & Halsey, T. C. Granular gravitational collapse and chute flow. *Europhys. Lett.* **60** (6), 931–937 (2002).

7. Lemaître, A. Origin of a repose angle: kinetics of rearrangement for granular materials. *Phys. Rev. Lett*. **89**, 064303 (2002).

8. Mohan L. S., Rao K. K. & Nott P. R. A frictional cosserat model for the slow shearing of granular materials. *J. Fluid Mech.* **457**, 377-409 (2002).

9. Josserand, C., Lagrée, P.-Y. & Lhuillier, D. Stationary shear flows of dense granular materials: a tentative continuum modelling, *Euro. Phys. J. E* **14**, 127-135 (2004).

10. Kumaran, V. Constitutive relations and linear stability of a sheared granular flow, *J. Fluid Mech.* **506**, 1-43 (2004).

11. Jaeger, H. M., Nagel, S. R. & Behringer, R. P. Granular solids, liquids, and gases. *Rev. Mod. Phys.* **68**, 1259–1273 (1996).

12. Goldhirsch, I. Rapid granular flows. *Annu. Rev. Fluid Mech.* **35**, 267–293 (2003).

13. Nedderman, R. M. *Static and kinematics of granular materials.* Cambridge University Press, Cambridge (1992).

14. GDR MiDi. On dense granular flows. *Euro. Phys. J. E* **14**, 341–365 (2004).

15. Da Cruz, F., Emam, S., Prochnow, M., Roux, J.-N. & Chevoir, F. Rheophysics of dense granular materials: Discrete simulation of plane shear flows. *Phys. Rev. E* **72**, 021309 (2005).

16. Iordanoff, I. & Khonsari, M. M. Granular lubrication: toward an understanding between kinetic and fluid regime. *ASME J. Tribol.* **126**, 137–145 (2004).

17. Pouliquen, O. & Forterre, Y. Friction law for dense granular flows: application to the motion of a mass down a rough inclined plane. *J. Fluid Mech.* **453**, 133–151 (2002).

18. Forterre, Y. & Pouliquen, O. Long-surface-wave instability in dense granular flows. *J. Fluid Mech.* **486**, 21–50 (2003).

19. Jop, P., Forterre, Y. & Pouliquen, O. Crucial role of sidewalls in dense granular flows: consequences for the rheology. *J. Fluid Mech.* **541**, 167–192 (2005).

20. Savage, S.B. The mechanics of rapid granular flows. *Adv. Appl. Mech.* **24**, 289–366 (1984).

21. Ancey, C., Coussot, P. & Evesque, P. A theoretical framework for very concentrated granular suspensions in a steady simple shear flow. *J. Rheol.* **43**, 1673–1699 (1999).

22. Pouliquen O. Scaling laws in granular flows down rough inclined planes. *Phys. Fluids* **11**, 542–548 (1999).

23. Drucker D.C. & Prager W. Soil mechanics and plastic analysis of limit design, *Quart. Applied Math.* **10**, 2 (1952).

24. Tanner, R. I. *Engineering Rheology*. Clarendon (1985).

25. Mueth, D. M. *et al.* Signature of granular microstructure in dense shear flows. *Nature* **406**, 385-389 (2000).

26. Fenistein, D. & Van Hecke, M. Wide shear zones in granular bulk flow. *Nature* **425**, 256 (2003).

27. Howell, D., Behringer, R. P. & Veje, C. Stress Fluctuations in a 2D Granular Couette Experiment: A Continuous Transition *Phys. Rev. Lett*. **82**, 5241-5244 (1999).

28. Daerr, A. & Douady, S. Two types of avalanche behaviour in granular media. *Nature* **399**, 241-243 (1999).

29. Malloggi, F., Lanuza, J., Andreotti, B. & Clément, E. Erosion waves: transverse instabilities and fingering. Submitted to *Phys. Rev. Lett.* Preprint at <http://arxiv.org/cond-mat/0507163> (2005).

30. Börzsönyi, T., Halsey, T. C. & Ecke, R. E. Two scenarios for avalanche dynamics in inclined granular layers. *Phys. Rev. Lett.* **94**, 208001 (2005).

Figure 1: Friction coefficient $\mu$ as a function of the dimensionless parameter $I$ ($\mu_s$=tan(20.9), $\mu_2$=tan(32.76) and $I_0$=0.279). Inset: definition of the pressure $P$, the shear stress $\tau$, and the shear rate $\dot{\gamma}$ in the simple plane shear configuration.

Figure 2: Experimental set-up of granular flows on a pile between rough sidewalls. The channel is 120 cm long and the width $W$ varies from 0.9cm (16.5 $d$) up to 28.9 cm (545 $d$).

Figure 3: Typical 3D velocity profile predicted by the rheology ($W$=142$d$, $\theta$ =22.6°, $Q/d^{3/2}g^{1/2}$=15.2). For sake of clarity only one quarter of the lines of the 71x80 computational grid is plotted.

Figure 4: Comparison of 3D simulations (lines) and experimental results (symbols) for different flow rates ($Q^*=Q/d^{3/2}g^{1/2}$) and different channel widths $W$ in the steady uniform regime. **a**, **b**, **c**, Free-surface velocity profiles for channel width $W$=16.5 $d$ (**a**), $W$=140 $d$ (**b**) and $W$=546 $d$ (**c**). **d**, Depths of the flowing layer across the channel for $W$=140 $d$. The experimental and computational flow rates are equal within 2.5%. The error bars represent the dispersion of the measurements for different experiments.

Figure 5: Scaling laws for the experimental measurements (symbols) compared to the predictions of the model (lines). Free-surface inclination tan($\theta$) (**a**), rescaled maximum free-surface velocity $\tilde{V}_{max}$ (**b**) and maximum flowing thickness $h_{max}/W$ (**c**), as a function of the rescaled flow rate $\tilde{Q}$. Data collapse onto single curves for all flow rates and all widths: $W$=16.5 $d$ (circles), $W$=55 $d$ (squares), $W$=140 $d$ (diamonds), $W$=546 $d$ (triangles). The grey circles (**a**,**b**) refer to measurements above the critical angle $\theta_2 = \arctan\mu_2$. In this case, no steady uniform flow is predicted by the model. Experimentally, we observe a transition to a collisional regime (see supplementary information and movies).

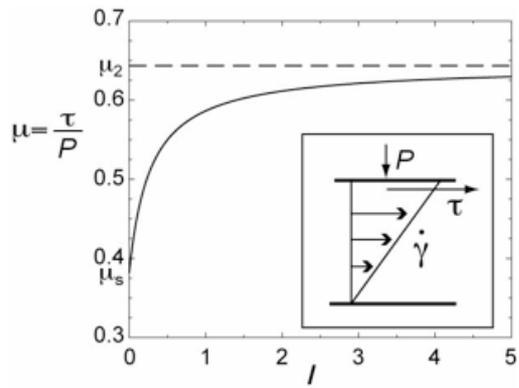

Figure 1.

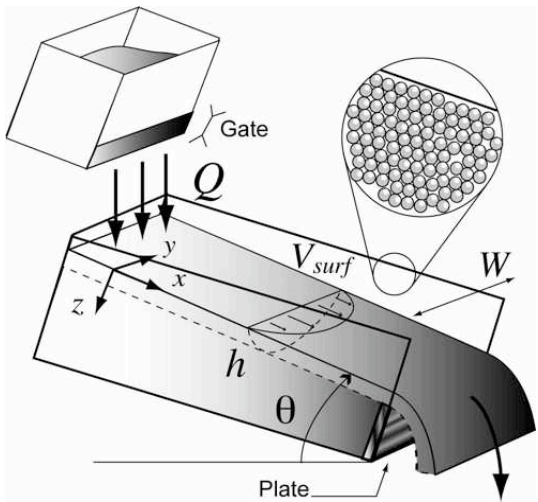

Figure 2

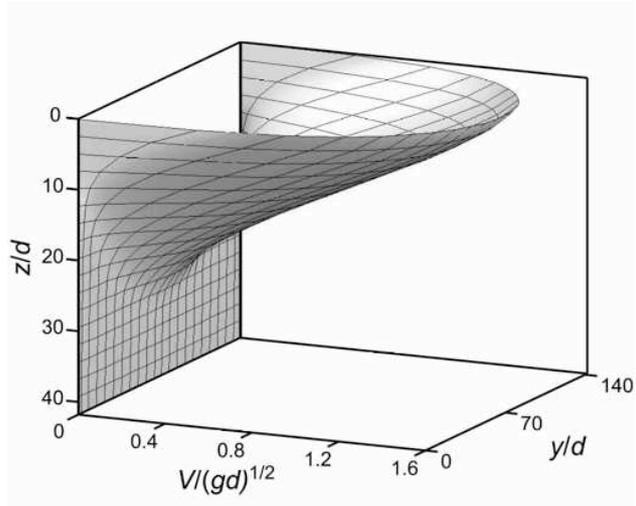

Figure 3

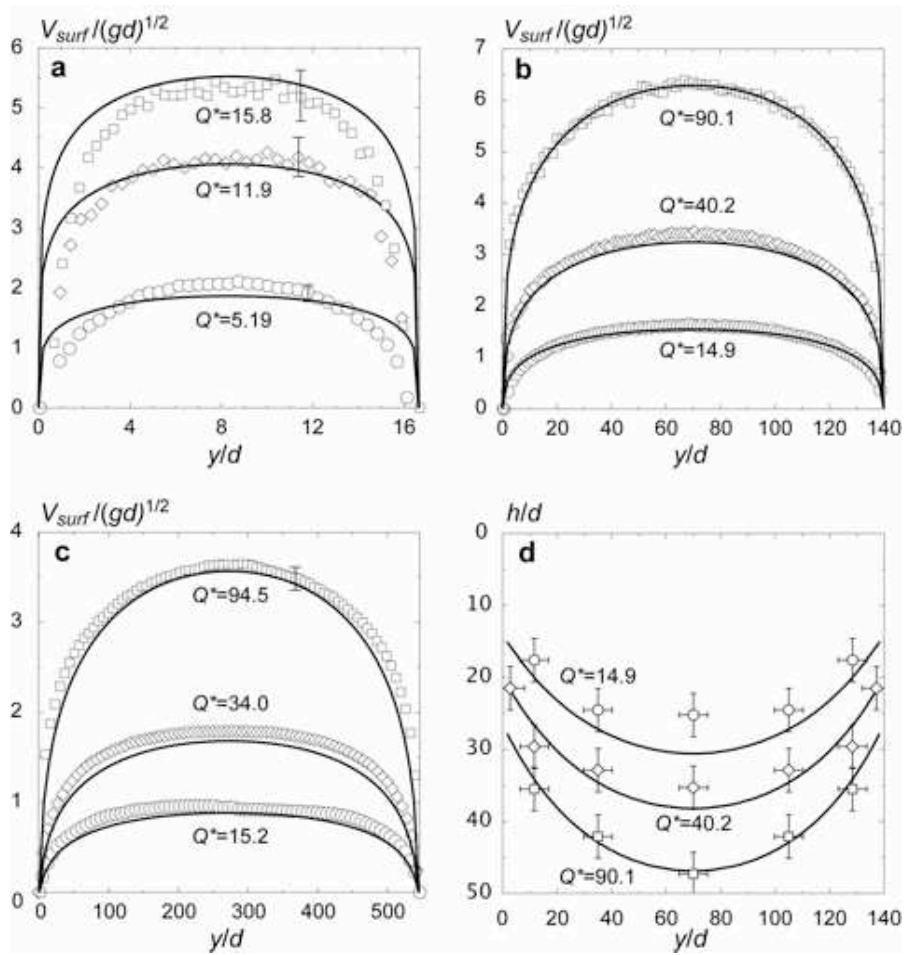

Figure 4

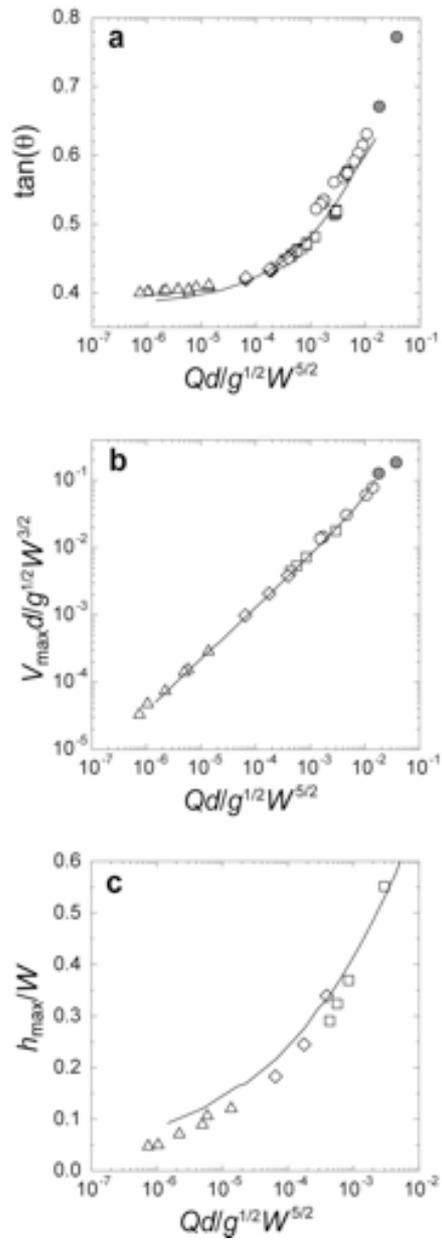

Figure 5